\documentclass[12pt]{article}
\usepackage{geometry}                
\usepackage{color}
\usepackage[pdftex]{graphicx}
\usepackage{subfigure}
\usepackage{wrapfig}
\usepackage{amssymb}
\usepackage{epstopdf}
\newcommand{\be}{\begin{equation}}       
\newcommand{\ee}{\end{equation}}       
\newcommand{\bear}{\begin{eqnarray}}       
\newcommand{\eear}{\end{eqnarray}}       
\newcommand{\ba}{\begin{array}}       
\newcommand{\ea}{\end{array}}

\newskip\humongous \humongous=0pt plus 1000pt minus 1000pt

\newif\ifdtup

\def\oldreffmt#1{\rlap{[#1]} \hbox to 2\parindent{}}

\def\figfmt#1{\rlap{Figure {#1}} \hbox to 1in{}}       
       
\newskip\humongous \humongous=0pt plus 1000pt minus 1000pt

\newif\ifdtup

\def\oldreffmt#1{\rlap{[#1]} \hbox to 2\parindent{}}

\def\figfmt#1{\rlap{Figure {#1}} \hbox to 1in{}}  
  
\def\ie{\hbox{\it i.e.}{}}        
\def\eg{\hbox{\it e.g.}{}}        
\def\etal{\hbox{\it et al.}}  
  
\def\beq{\begin{equation}}  
\def\eeq{\end{equation}}  
\def\bea{\begin{eqnarray}}  
\def\eea{\end{eqnarray}}

\def\bq{\begin{quote}}  
\def\eq{\end{quote}}

\def \etal {{\it et al.}\ } 

\def\cm2s{{\rm cm^{-2} s^{-1}}}
\relax  

\newdimen\tdim  
\tdim=\unitlength

\newskip\humongous \humongous=0pt plus 1000pt minus 1000pt

\newif\ifdtup

\def\oldreffmt#1{\rlap{[#1]} \hbox to 2\parindent{}}

\def\figfmt#1{\rlap{Figure {#1}} \hbox to 1in{}}  
  
\def\ie{\hbox{\it i.e.}{}}        
\def\eg{\hbox{\it e.g.}{}}        
\def\etal{\hbox{\it et al.}}  
  
\def\beq{\begin{equation}}  
\def\eeq{\end{equation}}  
\def\bea{\begin{eqnarray}}  
\def\eea{\end{eqnarray}}

\def\bq{\begin{quote}}  
\def\eq{\end{quote}}

\def \etal {{\it et al.}\ }  
\relax  

\newdimen\tdim  
\tdim=\unitlength

\begin{document}
\baselineskip=18pt \pagestyle{plain} \setcounter{page}{1}

\begin{flushright}       
Fermilab-CONF-13-284-T \\ 
July 16, 2013  
     \end{flushright}

\vspace*{1.2cm}

\begin{center}
{\Large \bf
The Case for a\\
Muon Collider Higgs Factory 
}
\vspace*{1.2cm}

{\normalsize 
  Yuri ~Alexahin$^{(1)}$,
  Charles~M.~Ankenbrandt$^{(2)}$,
  David~B.~Cline$^{(3)}$, 
  Alexander~Conway$^{(4)}$, 
  Mary~Anne~Cummings$^{(2)}$, 
  Vito~Di~Benedetto$^{(1)}$, 
  Estia~Eichten$^{(1)}$, 
  Jean-Pierre Delahaye$^{(5)}$,
  Corrado~Gatto$^{(6)}$,
  Benjamin~Grinstein$^{(7)}$,
  Jack~Gunion$^{(8)}$,  
  Tao~Han$^{(9)}$, 
  Gail~Hanson $^{(10)}$,  
  Christopher~T.~Hill$^{(1)}$, 
  Fedor~Ignatov$^{(11)}$,
  Rolland~P.~Johnson$^{(2)}$,
  Valeri~Lebedev$^{(1)}$,
  Leon M. Lederman$^{(1)}$,
  Ron~Lipton$^{(1)}$, 
  Zhen~Liu $^{(9)}$, 
  Tom~Markiewicz$^{(5)}$,
   Anna~Mazzacane$^{(1)}$,
   Nikolai~Mokhov$^{(1)}$,
   Sergei Nagaitsev $^{(1)}$, 
   David~Neuffer$^{(1)}$, 
   Mark~Palmer$^{(1)}$, 
   Milind~V.~Purohit$^{(12)}$,
   Rajendran Raja$^{(1)}$,
   Carlo Rubbia$^{(13)}$
   Sergei~Striganov$^{(1)}$,
   Don~Summers$^{(14)}$,  
   Nikolai~Terentiev$^{(15)}$,  
   Hans~Wenzel$^{(1)}$
                             } \\ [1cm]

{\small (1) {\it Fermilab, P.O. Box 500, Batavia, Illinois, 60510; } \\ }
{\small (2) {\it Muons Inc., Batavia, Illinois, 60510 } \\ }
{\small (3) {\it University of California, Los Angeles, California; }\\ }
{\small (4) {\it University of Chicago, Chicago, Illinois; }\\ }
{\small (5) {\it SLAC National Accelerator Laboratory, Palo Alto, California; }\\}
{\small (6) {\it INFN Naples, Universita Degli Studi di Napoli Federico II, Italia;}\\ }
{\small (7) {\it University of California, San Diego, California; }\\ }
{\small (8) {\it University of California, Davis, California; }\\ }
{\small (9) {\it University of Pittsburgh, Pittsburgh, Pennsylvania;}\\ }
{\small (10) {\it University of California, Riverside, California; }\\ }
{\small (11) {\it Budker Institute of Nuclear Physics, Russia;}\\ }
{\small (12) {\it University of  South Carolina,  Columbia, South Carolina; }\\ }
{\small (13) {\it CERN, Geneva, Switzerland;}\\ }
{\small (14) {\it University of Mississippi, Oxford, Miss.; }\\ }
{\small (15) {\it Carnegie Mellon University, Pittsburgh, Pennsylvania;}\\ }

\end{center}

\vspace*{0.1cm}

\newpage
                         
\section*{Introduction}

We propose the construction of a 
compact Muon Collider Higgs Factory. 
Such a machine can produce up to $\sim 14,000$ at $8\times 10^{31}$ cm$^{-2}$
sec$^{-1}$ clean Higgs events per year, enabling the most precise possible measurement of the mass, width and Higgs-Yukawa coupling constants (\eg , that of the muon).

A Muon Collider Higgs Factory is part of an evolutionary program 
beginning with aggressive R\&D on muon cooling, a possible neutrino factory such as $\nu$STORM, and the construction of Project-X with a rich program of precision physics addressing the $\sim 100$ TeV scale. This is followed by
the Muon Collider Higgs factory.  This program is upwardly scalable in energy
and can lead ultimately to the construction
of an energy frontier Muon Collider, reaching energy
scales in excess of $\sim 10$ TeV.
The Muon Collider Higgs Factory would utilize the intense proton beam 
from Project-X to produce, collect and cool muons, and by clever staging also use Project-X to accelerate the $\mu^+$ and $\mu^-$ bunches to $\sim 62.5$ GeV.
These would be placed in a $\sim 100$ m diameter storage ring (about the size
of the Fermilab Booster) and collide at single IP inside of dedicated detector.

The machine and detector issues at a Muon Collider 
have matured rapidly.
Complex timing can be deployed to suppress backgrounds.
A complete matrix for the machine now exists. Cooling has reached a conceptual stage at which  an R\&D program is required to establish
proof of principle of this component of the Muon Collider strategy. 
We are confident that this could be accomplished,
with sufficient funding, on a five-year time scale, leading directly to a 
design and preparation for construction of a Muon Collider Higgs Factory by the middle of the next decade.

For the 2013 Snowmass study we convened a workshop at UCLA. This 
included a detailed overview of the status of the Muon Collider and its
physics justification.
We did not focus on 6D cooling or many other machine issues that are
the subject of the MAP white paper \cite{MAPsnowmass}. 
Key conclusions of the UCLA Workshop were as follows \cite{key}:

\begin{itemize}

\item
The study of the Muon Collider Higgs Factory lattice indicates an attainable resolution of a few MeV, adequate to scan and measure the Higgs boson mass with a precision of $\sim 0.1$ MeV.
 
\item Precision measurement of the Higgs boson mass, width, Higgs-Yukawa coupling constants, etc., can reveal the nature of the Higgs boson and discriminate between various theoretical options.

\item  Preliminary studies to reduce the machine induced backgrounds using e.g. sophisticated timing and requiring tracks to eminate from the IP look very promising but will require further detailed studies including full simulation and reconstruction of physics events on top of background.

\item The Muon Collider Higgs Factory is the only proposed machine that can be upgraded to the multi-TeV scale with reasonable luminosity; a lack of significant beamstrahlung and therefore a narrow luminosity spectrum;  and power and cost effectiveness.  

\item The Muon Collider Higgs Factory is part of an evolutionary process, from Project-X to the multi-TeV scale energy frontier, offering a diverse and rich physics program along the way at both the intensity and energy frontiers.

\end{itemize}

 The concept of a Muon Collider was invented in the USSR in 1970's \cite{russ}.
Neuffer proposed a $90$ GeV Muon Collider ``Z-factory'' \cite{Neufferc} in 1979 and the muon storage ring for neutrino experiments \cite{Neuffern}.  Work in the USA ramped up in 1992 with the first dedicated workshop organized by UCLA
\cite{Thiessen}, and  the Muon Collider Higgs Factory concept emerged from this meeting \cite{Cline}.  During the 1990's, five workshops were organized by UCLA and two by BNL that served to
establish the feasibility of the Muon Collider concept.
In 1995 a Muon collider collaboration was formed with FNAL, BNL, LBNL and
several universities and regular annual collaboration meetings were held.
The muon storage ring neutrino factory concept became 
increasingly popular with the discovery of Neutrino Oscillations.
D. Cline and G. Hanson co-organized the muon
collider working groups at the 2001 Snowmass meeting. 
In 2010 the MAP program was formed and is now coordinating all Muon Collider
activity, directed by M. Palmer of Fermilab. 

There has been considerable recent progress on the physics studies of
the Higgs boson at a Muon Collider and other Higgs factories. The Higgs boson can be located and scanned within a year at luminosities of order $\sim 10^{32}$,
as detailed in the  2013 Snowmass  Muon Collider Higgs Factory white paper \cite{Higgs_whitepaper}.

Today there are only three realistic options for a lepton based
Higgs factory:

\begin{enumerate}

\item  An ILC type machine to produce the Higgs boson in associated production with a $Z^0$ requiring $10^{34}$ cm$^{-2}$ sec$^{-1}$ luminosity.
 
\item A compact circular Muon Collider Higgs Factory that produces the Higgs boson directly as  an  $s$-channel  resonance 
requiring $\sim 10^{32}$ cm$^{-2}$ sec$^{-1}$.

\item A large circular $e^+e^-$ collider, \ie, ``TLEP,''  of $\sim 80$ km circumference, also producing the Higgs boson in associated production with a $Z^0$
 requiring $\sim 10^{34}$ cm$^{-2}$ sec$^{-1}$ luminosity.

\end{enumerate}

In case (2)  the collider 
could be evolved up in energy to a multi-TeV Muon Collider while the 
core muon production, accumulation and cooling systems would remain almost the same.  
Likewise, a large circumference circular $e^+e^-$ collider as in case (3) could provide the tunnel for a 
(VLHC) at $\sim 100$ TeV proton-(anti)proton center-of-mass energy
addressing multi-TeV partonic energy scales. Such an evolution to the multi-TeV energy scales is not possible for option (1).

We quote some comments on the Muon Collider and its physics
potential from a recent
presentation of C. Rubbia \cite{Rubbia}:

\begin{itemize}       
 
\item  ``In a $\mu^+\mu^-$ collider, when compared to an $e^+ e^-$ collider, the direct Higgs boson production cross section is greatly enhanced since the $s$-channel coupling to a scalar is proportional to the lepton mass.''

\item ``Therefore the properties of the Higgs boson can be detailed over a larger fraction of model parameter space than at any other proposed accelerator method.''

\item ``A high energy  $\mu^+\mu^-$  collider is the only possible circular high energy lepton collider that can be situated within the CERN or FNAL sites.''

\item ``The unique feature of the direct production of a Higgs boson in the $s$-state is that the mass, total width and all partial widths can be directly measured with remarkable accuracy.''

\end{itemize}

Many more details are presented in the 2013 Snowmass Muon Collider 
and Muon Collider Higgs Factory white papers \cite{MAPsnowmass,Higgs_whitepaper}.

\end{document}